\documentclass[twocolumn]{paper}

\usepackage[width = 17cm]{geometry}
\usepackage{tikz}
\usetikzlibrary{arrows.meta, positioning}
\usepackage{graphicx}
\usepackage{subcaption}
\usepackage{amsmath}
\usepackage{amssymb}
\usepackage{amsfonts}
\usepackage{authblk}
\usepackage{braket}
\usepackage[english]{babel}
\usepackage[backend=biber, style=phys, eprint=true]{biblatex}
\usepackage{hyperref}
\usepackage{xcolor}

\title{Revisiting semiclassical scalar QED in 1+1 dimensions }

\author[1,2\thanks{\texttt{\url{santiago.sanz@dipc.org}}}]{S. Sanz-Wuhl}
\author[3\thanks{\texttt{\url{jochen.zahn@itp.uni-leipzig.de}}}]{J. Zahn}

\affil[1]{DIPC\textemdash Donostia International Physics Center, Paseo Manuel de Lardizábal 4, 20018 San Sebastián-Donostia, Spain.}
\affil[2]{Universidad de País Vasco/Euskal Herriko Unibertsitatea (UPV/EHU), 20018 San Sebastián-Donostia,  Spain.}
\affil[3]{Institut für Theoretische Physik, Universität Leipzig, Brüderstraße 16, 04103 Leipzig, Germany. }
\date{}                     

\begin{document}
\maketitle

\begin{abstract}
We study the backreaction of a charged scalar quantum field in the presence of two opposite charges placed at the boundaries of a finite one-dimensional region, with attention to boundary effects. 
We review, correct, and extend previous corresponding work of Ambjørn \& Wolfram \cite{ambjorn_properties_1983}.
Despite notable differences, our analysis confirms the mechanism, discussed by Ambjørn \& Wolfram, by which the incorporation of backreaction avoids certain instabilities. We also observe the interesting phenomenon of ``over-screening'', by which for high external charges an increase of the external charges leads to a decrease of the electric field between the two charges.

\end{abstract}

\section{Introduction}

Vacuum polarization is a phenomenon in quantum electrodynamics (QED) by which the vacuum of a charged field polarizes when subject to external electric fields, effectively behaving like a dielectric with $\epsilon>0$. This mechanism was one of the first QED effects studied \cite{dirac_discussion_1934,uehling_polarization_1935,heisenberg_consequences_1936}.
Observable consequences manifest e.g.~in corrections to atomic levels \cite{lamb_fine_1947} (cf. Ref.~\cite{mohr_qed_1998} for a review), while other effects e.g.~the Schwinger effect \cite{schwinger_gauge_1951} lie beyond current technological advances (cf. Ref.~\cite{fedotov_advances_2023}).

Defining vacuum polarization (or any other observable involving the product of fields or their derivatives at coinciding points) for generic backgrounds is not straightforward. Historically, \cite{dirac_discussion_1934} first defined it through the assumption that the short-distance behavior of the physically relevant states in the presence of strong background electric fields is the same as in their absence, up to smooth coefficients. In modern terminology, these states are called Hadamard, (cf. Ref.~\cite{fewster_necessity_2013}), and they are particularly interesting in the context of quantum field theory on curved spacetimes, where the same idea is used to define observables necessary for the study of e.g.~perturbative interacting field theory or semiclassical gravity. Detailed discussions on the definition of such observables can be found in Refs.~\cite{hollands_local_2001,hollands_quantum_2015}.

A further motivation to study vacuum polarization comes from semiclassical gravity. There, one
couples classical gravity to the expectation value of the stress-energy tensor, which similarly to the vacuum polarization needs to be properly defined through renormalization. 
But, to the best of our knowledge, solutions to semiclassical gravity have been confined to highly symmetric spacetimes, e.g.  \cite{sandersStaticSymmetricSolutions2022a,gottschalk_cosmological_2023}. It is thus desirable to dispense with symmetry assumptions, at least in toy models.

The (toy) model that we consider here consists of the charged scalar quantum field in 1+1 dimensional Minkowski spacetime, confined to a finite spatial interval and subject to an external electric field produced by charges at the boundaries of the interval. In particular, we consider the backreaction of the induced vacuum polarization on the classical electric field, i.e., study the semiclassical Maxwell-Klein-Gordon (MKG) system (an analog of the semiclassical Einstein equation).
This setup was already studied by Ambjørn \& Wolfram in \cite{ambjorn_properties_1983}, who in particular found that incorporation of backreaction cures instabilities which would otherwise be present.
However, as pointed out in \cite{schlemmer_current_2015}, the expression used in \cite{ambjorn_properties_1983} for the vacuum polarization $\rho$  cannot be derived from a manifestly gauge invariant prescription. 
The correct expression for $\rho$ was derived in \cite{wernersson_vacuum_2021}, and, as we will see, the resulting vacuum polarization in general differs qualitatively from the one obtained and used in \cite{ambjorn_properties_1983}. 

This motivates revisiting the results of Ambjørn \& Wolfram in \cite{ambjorn_properties_1983}, i.e., redoing their analysis but with the correct expression for the vacuum polarization $\rho$. We find that, even though the correct vacuum polarization differs substantially from the one obtained in \cite{ambjorn_properties_1983} (in particular for small external charges), the stabilization mechanism identified in \cite{ambjorn_properties_1983} persists. Additionally, we find the interesting phenomenon of ``over-screening'', by which, beyond a critical value, an increase of the external charges leads to a decrease of the electric field between the two charges.

The stabilization mechanism identified in \cite{ambjorn_properties_1983} (and confirmed in the present work) in the setting of scalar QED in 1+1 dimensions may also be relevant in the context of over-critical Coulomb potentials in QED in 3+1 dimensions \cite{Rafelski:1974rh}.

The layout of this article is as follows: Section 2 describes the setup, presents the semiclassical approximation and introduces a convenient gauge fixing. Sec.~\ref{sec:field-modes-and-vacuum-polarization} discusses the mode decomposition of the scalar field, and reviews the definition of vacuum polarization through a gauge invariant renormalization prescription, also comparing with the approach of \cite{ambjorn_properties_1983}. Sec.~\ref{sec:The-iterative-procedure} introduces an iterative procedure for solving the semiclassical MKG equations, and the results are shown in Sec.~\ref{sec:Results}, which are compared with the results of \cite{ambjorn_properties_1983}. We conclude with a summary and an outlook.

This article summarizes and extends the results of the M.Sc.~thesis of Santiago Sanz-Wuhl, written in 2025 at the Institut für Theoretische Physik, Universität Leipzig, under the supervision of Jochen Zahn. 
\section{Setup} 
\label{sec:Set-up}

In 1+1 dimensional spacetime, two charges $q, -q$
are set at $x^1=0, x^{1}=a$, respectively. Ignoring vacuum polarization effects,
this yields a constant electric field of strength $q$ pointing towards positive $x^{1}$ for $x^1\in[0, a]$.
We study stationary solutions to the semiclassical MKG equations, i.e.~the system of coupled differential equations
\begin{subequations}
    \begin{align}
        \left( D_\mu D^\mu + m^2 \right)\phi &= 0, 
        \label{eq:KG}\\
        \partial_\mu F^{\mu\nu} &= \left< j^\nu \right>,
        \label{eq:Maxwell}
    \end{align}
\end{subequations} in the region $x=(x^0, x^1) \in     \mathbb{R} \times [0,
a]$. Here, $D_\mu = \partial_\mu +ie A_\mu$ is the gauge covariant 
derivative, \begin{align}
    j^\nu(x) = ie \left((D^\nu \phi^*(x)) \phi (x)- \phi^*(x)
    D^\nu \phi(x) \right)
    \label{eq:charge-current-density}
\end{align}
the charge current density of the Klein-Gordon field, 
$\left< \mathcal{O}\right>$ is the renormalized vacuum expectation value of an observable $\mathcal{O},$ and the metric signature
is $(+,-)$.
We solve for $F_{\mu\nu}$ and (the mode decomposition of) the two-point function of the vacuum state.

In 1+1 dimensions, $A_1$ can be arbitrarily set to 0. This way, Eq.~\eqref{eq:Maxwell} reduces to the Poisson equation for the electrostatic potential $A_0$
\begin{align}
    \partial_1^2 A_0(x^1) = -\left< j^0 \right> \equiv -\rho, \label{eq:poisson}
\end{align} 
with $\rho$ the vacuum polarization, i.e.~the vacuum expectation value of the zeroth component of the charge current density.
Eq.~\eqref{eq:poisson} will be solved with the boundary conditions
\begin{align}
    \left. \partial_1
        A_0(x^1)\right|_{x^1=0, a} =  -q.
        \label{eq:A0-boundary-conditions}
\end{align}
This ensures that the electric field close enough to the boundary is not affected by vacuum polarization.
The boundary conditions \eqref{eq:A0-boundary-conditions} fix the solution to Eq.~\eqref{eq:poisson} up to an additive constant,
which is for convenience chosen so that $A_0(\frac{a}{2}) = 0$. With these boundary conditions and for a given $\rho(x^1)$, Eq.~\eqref{eq:poisson} is solved by
\begin{align}
        A_0(x^1) = -q\left(x^1 - \frac{a}{2} \right) -
        \int_{\frac{a}{2}}^{x^{1}} \int_{0}^{\chi} \rho(\xi) d \xi d \chi.
    \label{eq:A0-general} 
\end{align}

\section{Field Modes and Vacuum Polarization}
\label{sec:field-modes-and-vacuum-polarization}

In our gauge, the Klein-Gordon equation is of the form \begin{align}
    \left[ (\partial_0 + ie A_0 )^2 + \partial_1^2 - m^2
    \right]\phi = 0,
    \label{eq:TDKGE}
\end{align}
which we will study using Dirichlet boundary conditions (DBC) $\left.\phi \right|_{x^{1}=0,a} = 0$ or Neumann boundary conditions (NBC) $\left.\partial_{1} \phi\right|_{x^1=0,a }=0$. More general boundary conditions may be considered, but these are outside the scope of this work.
Using the mode ansatz $\phi_n(x) = \phi_n (x^{1}) e^{-i\Omega_n x^0}$  the PDE is split into ordinary differential equations for each mode $\phi_n (x^{1})$ of the scalar field
\begin{align}
    \left[ (\Omega_n - e A_0 )^2 + \partial_{1}^2 -  m^2 \right]\phi_n
    = 0.
    \label{eq:TIKGE}
\end{align}
It is convenient to introduce the dimensionless parameters  
\begin{align} 
z & = a^{-1} x^1, &
\lambda & = e q a^2, &
\varepsilon & = a e, &
\omega_{n} & = a \Omega_{n}, 
\end{align}
and to treat $\phi$ as a function of $z$ (instead of $x^1$). While $\lambda$ parameterizes the strength of the external charges, $\varepsilon$ characterizes the strength of the coupling between electric and scalar field.

 
In terms of the above dimensionless parameters, Eq.~\eqref{eq:TIKGE} takes the form \begin{align}
    \left[ (\omega_n - \varepsilon A_0 )^2 + \frac{d^2}{dz^2} - a^2
    m^2\right] \phi_n = 0.  \label{eq:mode-equation}
\end{align}
The modes $\phi_n$ are normalized with respect to the symplectic
norm
\begin{align}
    (\phi_n, \phi_m) = i\int_{0}^{1} (\phi_n^* \pi_m^* - \pi_n \phi_m)dz,
\end{align}
with $\pi_n = (D_0 \phi_n)^*$ the canonical conjugate of
$\phi_n$.

The general solution to Eq.~\eqref{eq:TDKGE} is thus decomposed in
modes \begin{align}
    \phi(t, z) = \sum_{n>0}^{} a_n \phi_n(z) e^{-i\omega_n t} +
    \sum_{n<0}^{} b^\dagger_n \phi_n(z) e^{-i\omega_n t}.
    \label{eq:mode-decomposition}
\end{align} Positive (negative) values of $n$ are identified with modes of positive
(negative) symplectic norm. $a_n$ and $b_n$ are the ladder operators for the positive and
negative norm solutions, respectively, obeying canonical commutation relations
$
    [a_n, a_m^\dagger ] = [b_n, b_m^\dagger ] =\delta_{nm},
$
and all other commutators vanishing. The ladder operators define
the vacuum state $\ket{0}$ as that for which $a_n \ket{0} = b_n\ket{0} = 0$. Note that due to the chosen gauge, $\omega_n = -\omega_{-n},$ and $b_n^\dagger$  is the creation operator of the $n$-th negative frequency mode. 

\subsection{Vacuum Polarization}
\label{sec:vacuum-polarization}
Quantities such as vacuum polarization $\rho$, involving products of (derivatives of) fields at coinciding points, are a priori ill-defined and need to be properly defined through a local and gauge invariant renormalization scheme. Such a scheme is the Hadamard point-split scheme, proposed in \cite{dirac_discussion_1934} in the context of QED in external electromagnetic fields and ``rediscovered'' in the 1970s in the context of QFT in curved spacetimes \cite{adlerRegularizationStressenergyTensor1977, waldTraceAnomalyConformally1978}. For a modern formulation, we refer to \cite{hollands_local_2001, hollands_quantum_2015}. The scheme is applicable whenever the state is of Hadamard form, which will in particular be the case for the ground states we are considering \cite{wrochna_quantum_2012}. For a general discussion on the Hadamard condition, we refer to \cite{fewster_necessity_2013}. 
The two-point function
\begin{equation}
    w_{\Omega}^{\phi \phi^*}(x, x') = \braket{\phi (x)\phi^* (x')}_\Omega
\end{equation}
of the charged scalar field $\phi$ in
a Hadamard state $\Omega$  is of the form
\begin{align}
    w_{\Omega}^{\phi\phi^*}(x, x') 
    \equiv H^{\phi \phi^*}(x, x')  + R^{\phi \phi^*}_{\Omega} (x, x').
    \label{eq:hadamard-condition}
\end{align}
Here,
\begin{equation}
    H^{\phi\phi^*}(x,x') = -\frac{V(x,x')}{4\pi} \ln \frac{- (x-x')^2 + i 0_+ (t-t')}{\Lambda^2}
    \label{eq:hadamard-parametrix}
\end{equation}
is the Hadamard parametrix of the scalar field in 1+1 spacetime dimensions and $R_{\Omega}^{\phi\phi^*}$
a state-dependent smooth function. $\Lambda$ is an arbitrary length scale and
the smooth coefficient $V$, up to the order relevant for vacuum polarization reads \cite{schlemmer_current_2015}
\begin{align}
    V(x, x') &= \exp\left[-i\varepsilon \int_{0}^{1} A_{\mu}(x' +
    s(x-x') ) (x - x')^{\mu}ds \right] \nonumber \\&+ \mathcal{O}\left((x-x')^2\right).
\end{align} 
The Hadamard condition allows us to define expectation values of
local observables non-linear in the fields as a limit of coinciding points:
\begin{align}
\begin{split}
         \braket{D_\alpha \phi (x) D^{*}_\beta \phi^*(x)}_\Omega & \\= \lim_{x' \to x}
    D_\alpha D^{*}_\beta{}' &(w_\Omega^{\phi \phi^*}(x, x')- H^{\phi\phi^*}(x,
    x')),
\end{split}
    \label{eq:point-splitting}
\end{align}
with $\alpha, \beta$ symmetrized multiindices. 


Eqs.~\eqref{eq:point-splitting}, \eqref{eq:charge-current-density} together with  the mode decomposition \eqref{eq:mode-decomposition}  result in the expression for $\rho$ calculated in \cite{wernersson_vacuum_2021} 
\begin{align}
   \label{eq:vacuum-polarization-long}
        \rho =& \varepsilon \lim_{\tau \to 0}  \Big[\sum_{n>0}^{}(\omega_n - \varepsilon A_0) \lvert
    \phi_n\rvert^2 e^{i\omega_n ( \tau + i 0_+) }\\ 
    &+\sum_{n<0}^{} (\omega_n - \varepsilon
    A_0) \lvert \phi_n\rvert^2 e^{-i\omega_n ( \tau + i 0_+) } \Big]+ \frac{\varepsilon^2}{\pi}
    A_0 . \nonumber
\end{align}
A discussion of how to evaluate this expression in practice can be found in \cite{wernersson_vacuum_2021}.
An example of $\rho$ corresponding to the potential $\varepsilon A_0(z) = -\lambda(z- \frac{1}{2})$ with $\lambda=1$ is shown in Fig.~\ref{fig:hambjorn}.

To measure the \textit{strength} of the vacuum polarization, the \textit{induced charge}
\begin{align}
    Q_I = \int_{0}^{\frac{1}{2}} \rho(z) dz
    \label{eq:polarized charge}
\end{align}
is defined, which is expected to be negative. This charge partially screens the charge at $z=0$
and the total electric field at $z=\frac{1}{2}$ is $\lambda + Q_I$.

\subsection{External Field Approximation and Critical Field Strength}
\label{subsec:External-Field-Approximation}

The external field approximation ignores the contribution of the backreaction
of the charged scalar field to the background electric field. This
approximation highlights the potential instabilities of the system, which are remedied by the incorporation of backreaction.
Eq.~\eqref{eq:mode-equation} is solved with 
$\varepsilon A_0(z) = -\lambda (z - \frac{1}{2})$, and takes the form
\begin{figure}[t]
    \centering
        \includegraphics[width=0.45\textwidth]{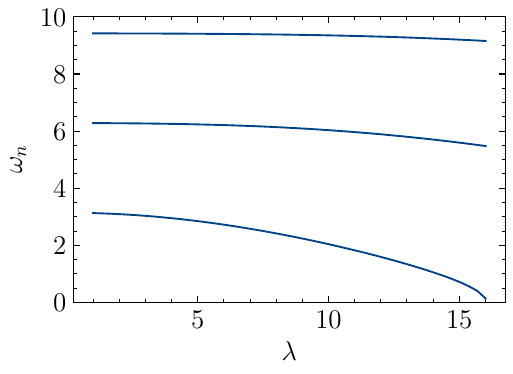}
        \caption{Energies of the three first modes of the massless
    Klein-Gordon field subject to DBC in the external field approximation
    as functions of  $\lambda.$ $\omega_1\to0$ as $\lambda\to\lambda_c \approx 16.05$ defines the critical field strength for these mass and boundary conditions configurations.}
    \label{fig:eigenvalues-external-field-approximation-kg}
\end{figure}
\begin{figure}[t]
    \centering
    \includegraphics[width=0.45\textwidth]{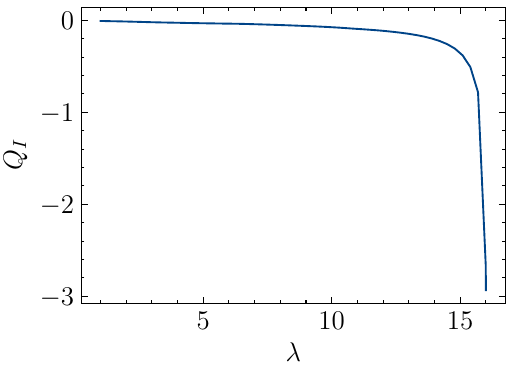}
    \caption{Corresponding induced charge $Q_I$ in Fig.~\ref{fig:eigenvalues-external-field-approximation-kg} as a function of $\lambda$.}
    \label{fig:QI-eigenvalues-external-field-approximation}
\end{figure}
\begin{align}
    \left[ \left(\omega_n + \lambda \left(z-\frac{1}{2}\right) \right)^2 +
    \frac{d {^2}}{d {z^2}} - a^2m^2
\right] \phi_n = 0.  
\label{eq:external-field-approximation-kg}
\end{align}
For each value of $\lambda$, the modes $\phi_n$ and their energies  $\omega_n$  
can be obtained from the boundary conditions.
Fig.~\ref{fig:eigenvalues-external-field-approximation-kg} displays the energies
$\omega_1, \omega_2, \omega_3$ of the first three modes  of the
massless scalar field with DBC as functions of the
background field strength $\lambda$. 
The value at which $\omega_1$ goes to $0$ defines the critical field strength $\lambda_c=16.05039(5)$.  Fig.~\ref{fig:QI-eigenvalues-external-field-approximation} displays the divergent behavior of  $Q_I(\lambda)$ for this configuration as $\lambda\to\lambda_c$. $\lambda_c$ takes different values for different boundary conditions and masses. For $\lambda \geq \lambda_c$, the first mode is no longer symplectically normalizable, so that no ground state for the scalar field exists (for $\lambda > \lambda_c$, the frequency $\omega_1$ becomes imaginary, indicating an instability already at the classical level).

%

\subsection{Mode sum formula}
\label{sec:mode-sum-formula}
Vacuum polarization in \cite{ambjorn_properties_1983} is defined through a so-called mode sum formula. It pairs the charge density associated to each mode as 
\begin{align}
    \rho^\text{MS}_n = \varepsilon \left[(\omega_n-\varepsilon A_0)\lvert \phi_n \rvert^2  - (\omega_n+\varepsilon A_0)\lvert \phi_{-n} \rvert^2\right]
    \label{eq:mode-charge-density}
\end{align}
to define the vacuum polarization 
\begin{align}
    \label{eq:mode-sum}    
        \rho^\text{MS} &= \sum_{n>0}^{} \rho^\text{MS}_n \\&= \varepsilon\sum_{n>0}^{}\big[(\omega_n-\varepsilon A_0)\lvert \phi_n \rvert^2
    - ( \omega_n +\varepsilon A_0) \lvert\phi_{-n}\rvert^2
    \big], \nonumber
\end{align}
where the superscript MS stands for a ``mode sum'' prescription.
In practice, the difference to the correct prescription~\eqref{eq:vacuum-polarization-long} is the missing term $\frac{\varepsilon^2}{\pi} A_0$, which originates from the Hadamard point-split (but see the discussion in \cite{wernersson_vacuum_2021} for the significance of the limit $\tau \to 0$ in \eqref{eq:vacuum-polarization-long}).
The above expression cannot be derived from a manifestly gauge invariant renormalization scheme and relies on the specific pairing of modes to be well-defined. 
Furthermore, in the iterative procedure incorporating backreaction (discussed in more detail below),~\cite{ambjorn_properties_1983} truncates the mode sum to the first mode, i.e., considers $\rho^\text{MS}_1$.

\begin{figure}
    \centering
    \includegraphics[width=0.45\textwidth]{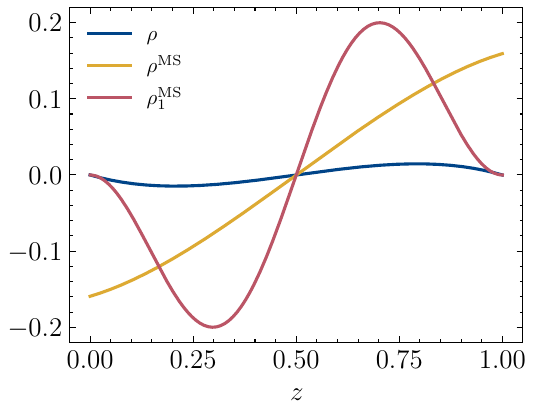}
    \caption{Three different $\rho$ for the same massless scalar field subject to DBC resulting from\linebreak$\varepsilon A_0 = -(z - \frac{1}{2})$ : In blue, the correct prescription in Eq.~\eqref{eq:vacuum-polarization-long} used throughout this paper. In orange, $\rho$ calculated through the mode sum \eqref{eq:mode-sum}. In red, $\rho$ defined through the mode sum formula, truncated to the first mode. }
    \label{fig:hambjorn}
\end{figure}

Fig.~\ref{fig:hambjorn} shows the charge densities arising from the different $\rho$ prescriptions, for the massless scalar field subject to DBC. The same $\phi_n$, $\omega_n$ and  \linebreak$\varepsilon A_0(z) = -\lambda(z-\frac{1}{2})$, the external field approximation with $\lambda=1$, have been used for the three curves. In blue, the correct vacuum polarization \eqref{eq:vacuum-polarization-long} (used throughout this paper), in orange the vacuum polarization \eqref{eq:mode-sum} from the mode sum prescription and in red its truncation $\rho^{\text{MS}}_1$ to the first mode (used for the study of backreaction in \cite{ambjorn_properties_1983}). 

We observe that $\rho$ differs qualitatively and quantitatively from the mode sum result $\rho^{\text{MS}}$ and its truncation $\rho^{\text{MS}}_1$, in that it is substantially weaker and vanishes at the boundary (as might be expected for Dirichlet boundary conditions). It is thus a priori unclear whether the stabilization mechanism identified in \cite{ambjorn_properties_1983} based on $\rho^{\text{MS}}_1$ persists when using the correct vacuum polarization \eqref{eq:vacuum-polarization-long}.

\section{Self-consistent fields}
\label{sec:The-iterative-procedure}

As discussed above, the polarization of the vacuum induces a charge density $\rho$, which modifies the background field as in Eq.~\eqref{eq:A0-general}. This modified background field can be used in Eq.~\eqref{eq:mode-equation-iterative} to calculate new mode solutions, that will lead to a different vacuum polarization. This iterative procedure, schematized in Fig.~\ref{fig:iterative-procedure}, is performed repeatedly, and we study its convergence.

Concretely, for each value of $\lambda$ (parameterizing the strength of the external charge $q$) and at each iteration $\kappa$, we find normalized $(\phi^\kappa_{n, \lambda}, \omega^\kappa_{n, \lambda})$, and corresponding $\rho^\kappa_\lambda$, $\varepsilon A^\kappa_{0, \lambda}$ solving
\begin{subequations}
\label{eq:iterative-procedure}
\begin{align}
    \frac{d {^2}\phi_{n, \lambda}^{\kappa}}{d {z^2}} {}  &= 
    -\left[ (\omega_{n, \lambda}^\kappa - \varepsilon A^{\kappa -1 }_{0, \lambda} )^2 
     - a^2 m^2\right] \phi_{n, \lambda}^{\kappa}, 
    \label{eq:mode-equation-iterative}
    \\
        \rho_\lambda^{\kappa} &= \varepsilon\lim_{\tau \to 0} \Big[\sum_{n>0}^{} (\omega_{n,
        \lambda}^\kappa - \varepsilon A_{0, \lambda}^{\kappa-1} )\lvert \phi_{n,
        \lambda}^\kappa\rvert^2 e^{i\omega^\kappa_{n, \lambda} (\tau + i 0_+)} \nonumber\\
        &+\sum_{n<0}^{} (\omega^{\kappa}_{n, \lambda} - \varepsilon
        A_{0, \lambda}^{\kappa-1}) \lvert \phi^\kappa_{n, \lambda}\rvert^2
    e^{-i\omega^{\kappa}_{n, \lambda} (\tau + i 0_+)}\Big] \nonumber \\
\label{eq:vacuum-polarization-iterative}
    &+
    \frac{\varepsilon^2}{\pi}    A_{0, \lambda}^{\kappa-1} , \\
        \label{eq:A0-general-iterative}
        \varepsilon A^{\kappa }_{0, \lambda} &= c \varepsilon A_{0, \lambda}^{\kappa-1} \\&+ \left(1-c
        \right)\left[ -\lambda\left(z - \frac{1}{2} \right) - \varepsilon
            \int_{\frac{1}{2}}^{z} \int_{0}^{\chi} \rho_\lambda^{\kappa}(\xi) d\xi d\chi
             \right] . \nonumber
\end{align}
\end{subequations}
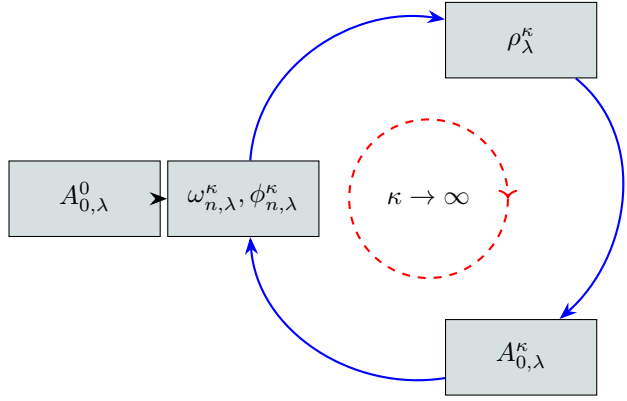
\begin{figure}[t]
\begin{center}
\begin{tikzpicture}[
    scale=0.7,
    box/.style = {draw, rectangle, minimum width=2cm, minimum height=1cm, fill={rgb:red,1;green,2;blue,2}, fill opacity=0.2, text opacity = 1},
    arrow/.style = {->, thick, >=Stealth},
    curved arrow/.style = {->, thick, >=Stealth, bend left}
]
\node[box] (A0) at (0, 0) {$A_{0,\lambda}^{0}$};
\node[box] (phi) at (3, 0) {$\omega_{n, \lambda}^\kappa, \phi_{n, \lambda}^\kappa$};
\node[box] (rho) at (8.25, 3) {$\rho_\lambda^\kappa$};
\node[box] (tildeA0) at (8.25, -3) {$A_{0,\lambda}^{\kappa}$};
\node (kappa) at (6.5, 0) {$\kappa \to \infty$};

\draw[arrow] (A0) -- (phi);

\draw[curved arrow, blue] (phi) to[out=50, in=135] (rho);
\draw[curved arrow, blue] (rho) to[out=55, in=135] (tildeA0);
\draw[curved arrow, blue] (tildeA0) to[out=45, in=130] (phi);


\draw[->, thick, dashed, red] (8, 0) arc[start angle=360, end angle=0, radius=1.5cm];

\end{tikzpicture}
\end{center}
\caption{Schematic of the iterative procedure for fixed $\lambda$.}%
\label{fig:iterative-procedure}
\end{figure}

This procedure has two main differences from the one used in \cite{ambjorn_properties_1983}: the expression for the vacuum polarization, and the introduction of a ``damping parameter'' $c \in [0,1)$, which will be discussed below.

Interpreting Eq.~\eqref{eq:A0-general-iterative} as an update map on $A_{0,\lambda}^{\kappa-1}$, we can think of the iterative procedure as an infinite-dimensional fixed-point problem. That is, for each $\lambda$, we seek $A_{0,\lambda}$ such that
\begin{align}
A_{0,\lambda} = f_{\lambda, c}(A_{0,\lambda}),
\label{eq:fixed-point-problem}
\end{align}
with $f_{\lambda, c}$ given by the composition of steps \eqref{eq:mode-equation-iterative}, \eqref{eq:vacuum-polarization-iterative} and \eqref{eq:A0-general-iterative}.
The fixed points of $f_{\lambda, c}$, denominated ``self-consistent solutions", are looked for by studying the limit $A_{0,\lambda}=\lim_{\kappa\to \infty}A_{0,\lambda}^{\kappa}$, where 
\begin{align} 
    A_{0,\lambda}^{\kappa + 1} = f_{\lambda, c}(A_{0,\lambda}^{\kappa}) =
    f_{\lambda, c}^{\kappa}\left(A_{0, \lambda}^{0} \right), 
    \label{eq:self-consistent-limit}
\end{align}
    with
\begin{equation}
    f_{\lambda, c}^\kappa =
    \underbrace{f_{\lambda, c} \circ ... \circ f_{\lambda, c}}_{\kappa \textrm{ times}}.
\end{equation}    
In a Banach space setting, by Banach's fixed point theorem \cite{banachOperationsDansEnsembles1922} the limit $A_{0, \lambda}$ will exist if the map $f_{\lambda,c}$ is a contraction, which is not ensured a priori. If this limit exists, we call the potential and the corresponding quantities ``self-consistent''.

For low enough $\lambda\ll \lambda_c$, the initial potential  $A_{0,
\lambda}^0$  can be taken to be the external field approximation  $\varepsilon  A_{0,
\lambda}^0(z) = -\lambda \left(z- \frac{1}{2}\right)$. 
This is clearly not a valid initial condition for $\lambda \geq \lambda_c$, as the evaluation of the corresponding vacuum polarization fails due to the instability discussed above. Hence, a better choice for the initial potential $A_{0, \lambda}^0$ is needed. But even with such a better choice, convergence of the iterative procedure is not guaranteed. In particular, it can happen that an update $A_{0,\lambda}^{\kappa+1} = f_{\lambda, c}(A_{0,\lambda}^{\kappa})$ ``overshoots'', leading to oscillatory or even run-away behavior, as exemplified in Appendix~\ref{sec:convergence}. Such an overshooting can be ``damped'' by choosing $c > 0$ in Eq.~\eqref{eq:A0-general-iterative}, which reduces the weight of the update (the damping is enhanced as $c \to 1$).
A numerical continuation method is used to traverse the $\lambda$-parameter:
\begin{enumerate}
    \item Assume a self-consistent $A_{0,\lambda}^\infty$ is known.
    \item Choose an increment $\Delta \lambda>0$, which defines a candidate $\lambda_{\text{new}} = \lambda + \Delta \lambda$.
    \item Perform the iterative procedure with the initial condition
    \begin{equation}
        A_{0, \lambda_{\text{new}}}^0 = A_{0, \lambda}^\infty.
    \end{equation}
    \item If this sequence converges, store the $\lambda$ and self-consistent fields, and proceed from step 1 with the newly calculated $A_0$. If it fails, reduce $\Delta \lambda$, and proceed from step 2.  
\end{enumerate}

\section{Results}%
\label{sec:Results}

\begin{figure}[p]
    \centering
        \begin{subfigure}{0.5\textwidth}
            \includegraphics[width=\textwidth]{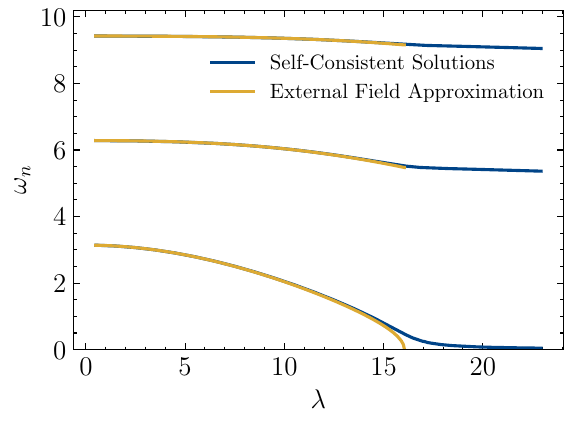}
        \end{subfigure}
        
                \begin{subfigure}{0.5\textwidth}
            \includegraphics[width=\textwidth]{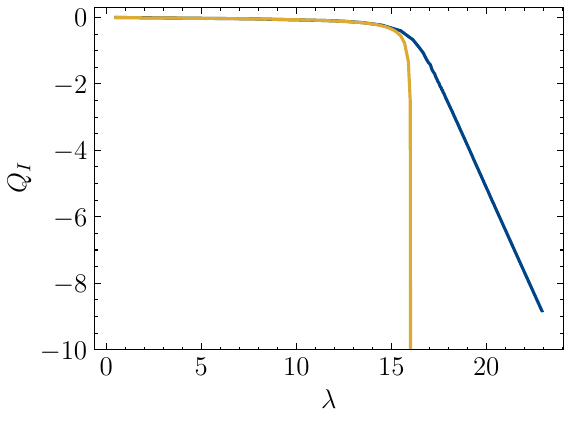}
        \end{subfigure}
    \begin{subfigure}{0.5\textwidth}
            \includegraphics[width=\textwidth]{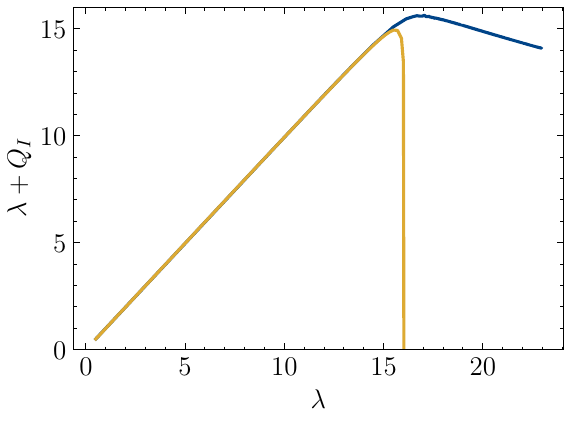}
        \end{subfigure}
        \caption{
            In blue, the self-consistent solutions to the coupled semiclassical Klein-Gordon-Maxwell system for a massless scalar field subject to DBC. In orange, for reference, the external field approximation is given. Top panel: Energies of the first three modes as functions of $\lambda$. Middle panel: Corresponding $Q_I$ for each $\lambda$. Bottom panel: Total electric field $\lambda+Q_I$ due to the screening of the vacuum at $z=\frac{1}{2}$. 
        }
        \label{fig:main-results-dirichlet}
\end{figure}

\begin{figure}[p]
    \centering
        \begin{subfigure}{0.5\textwidth}
            \includegraphics[width=\textwidth]{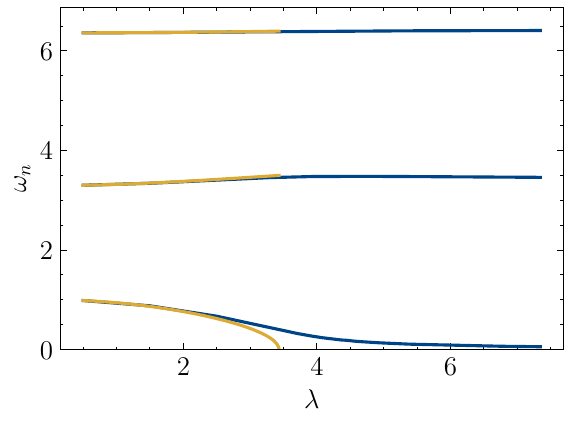}
        \end{subfigure}
                \begin{subfigure}{0.5\textwidth}
            \includegraphics[width=\textwidth]{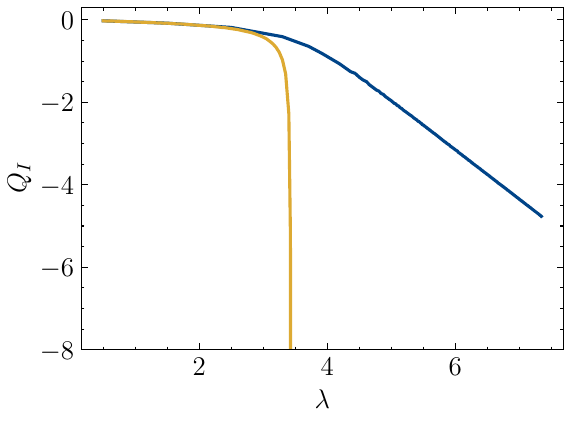}
        \end{subfigure}
            \begin{subfigure}{0.5\textwidth}
            \includegraphics[width=\textwidth]{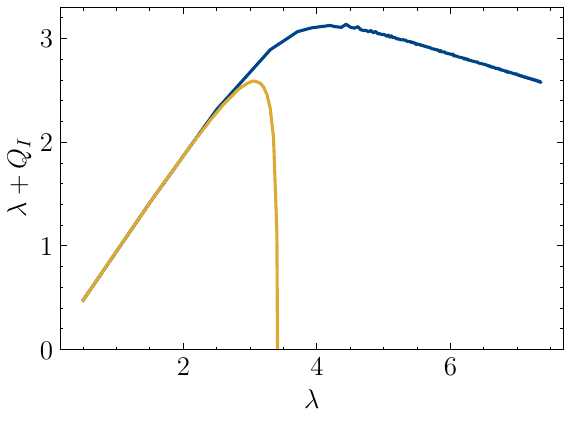}
        \end{subfigure}
        \caption{Following the format in Fig.~\ref{fig:main-results-dirichlet}, the results for the massive ($a m = 1$) scalar field subject to NBC. Further details in main text.}
        \label{fig:main-results-neumann}
\end{figure}

Using the iterative procedure laid out above we focus on two particular cases: The massless scalar field subject to DBC, and  $a m=1$ with
the scalar field subject to NBC. We do not study the massless field subject to NBC, since under these boundary conditions, $\lambda_c=0$ for $m=0$, as pointed out in Ref.~\cite{wernersson_vacuum_2021}. We work with $\varepsilon = 1$, to facilitate the comparison with \cite{ambjorn_properties_1983}.

Figs.~\ref{fig:main-results-dirichlet} and \ref{fig:main-results-neumann}  display the dynamics of the KGM system in the semiclassical approximation. The corresponding quantities in the external field approximation are given for reference in orange. Notice the different $\lambda$ scales for DBC and NBC. The system is studied through the energies of the first modes (top panel), the induced charge $Q_I$ (middle panel) and the total electric field $\lambda+Q_I$ at the midpoint $z=0.5$ (bottom panel). We observe in both cases how considering backreaction avoids the instabilities reflected in $\omega_1 \to 0$ and $Q_I\to-\infty$ as $\lambda \to \lambda_c$ in the external field approximation. We also point out how the rapid growth of $|Q_I|$ as $\lambda$ crosses $\lambda_c$ keeps $\lambda + Q_I$ below $\lambda_c$, which effectively cures the instabilities appearing in the external field approximation. Surprisingly, upon further increasing $\lambda$, $\lambda+Q_I$ decreases instead of stabilizing, a phenomenon one might term ``over-screening''.

One might also notice the almost identical behavior of the self-consistent solutions to the
external field approximation in the weak-field regime $\lambda \ll {\lambda_c}$, which seems to be independent of mass and boundary conditions.

\begin{figure}[t]
    \begin{center}
        \includegraphics[width=0.47\textwidth]{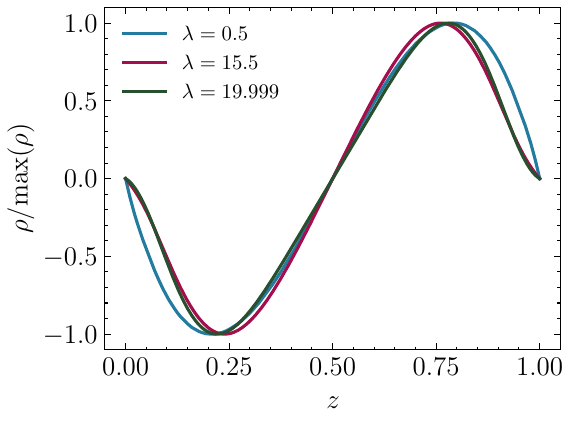}
    \end{center}
    \caption{The self-consistent vacuum polarization for varying $\lambda$ for the massless field subject to DBC. For visualization purposes, each curve is normalized with respect to its maximum value.}
    \label{fig:rhoEvolution-dirichlet}
\end{figure}

\begin{figure}[t]
    \begin{center}
        \includegraphics[width=0.47\textwidth]{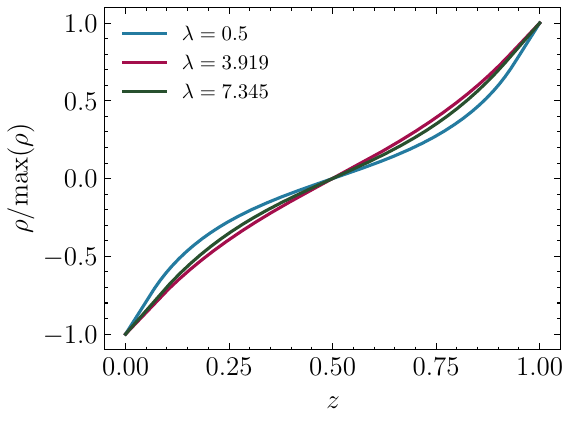}
    \end{center}
    \caption{Same as Fig.~\ref{fig:rhoEvolution-dirichlet} for the $a m=1$ field subject to NBC.}
    \label{fig:rhoEvolution-neumann}

\end{figure}

For completeness, the self-consistent vacuum polarizations for different values of $\lambda$ are displayed in Figs.~\ref{fig:rhoEvolution-dirichlet} and \ref{fig:rhoEvolution-neumann} corresponding to the results in Figs.~\ref{fig:main-results-dirichlet} and~\ref{fig:main-results-neumann}, respectively.


\subsection{Mode sum formula}

\begin{figure}
    \centering
    \includegraphics[width=0.45\textwidth]{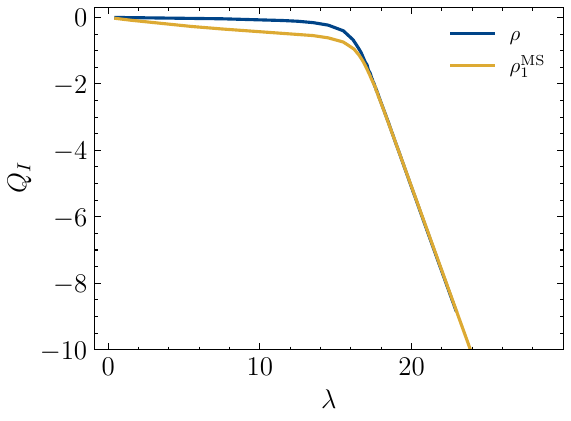}
    \caption{$Q_I$ of the self-consistent solutions for a massless field subject to DBC resulting from the two prescriptions of vacuum polarization described through the article.}
    \label{fig:hambjorn-iterative}
\end{figure}
 
We compare the results obtained in this work with those of \cite{ambjorn_properties_1983}.
Fig.~\ref{fig:hambjorn-iterative} displays $Q_I$ in the iterative procedure as a function of $\lambda$ for two different prescriptions for vacuum polarization: in blue, the correct prescription $\rho$ used throughout this paper, and in orange $\rho^\text{MS}_1$ as discussed in Sec.~\ref{sec:mode-sum-formula}. The two prescriptions exhibit markedly different behavior in the weak-field regime, where the truncated mode sum yields a noticeably stronger vacuum polarization. In the strong-field regime, however, their behavior is identical. This can be understood as a consequence of the low-energy modes being the most sensitive to the background field, and thus the ones that contribute most to vacuum polarization.
\section{Discussion and outlook}
\label{sec:discussion}

We revisited the work of Ambjørn \& Wolfram~\cite{ambjorn_properties_1983}, studying the effect of backreaction of a charged scalar quantum field in 1+1 dimensional spacetime, but with the correct expression for the vacuum polarization. We found that, despite qualitative and quantitative differences in the corresponding vacuum polarizations, the mechanism, identified by Ambjørn \& Wolfram, through which backreaction avoids certain instabilities, also applies to the correct vacuum polarization. In fact, in the strong field regime, our results are essentially indistinguishable from those obtained with the method of \cite{ambjorn_properties_1983}.

In the strong field regime, we also find the interesting effect of ``over-screening'', leading to a decrease of the electric field between the two charges with increasing external charge, as seen in the bottom plots of Figs.~\ref{fig:main-results-dirichlet} and~\ref{fig:main-results-neumann}.

Several further directions are worth exploring. As an extension of the setup, one could drop the boundary conditions imposed at the two external charges, leading to a continuous spectrum (and the need to work at non-vanishing mass). Technically considerably more challenging would be the study of the potential removal of instabilities through backreaction in overcritical potentials in QED in 3+1 dimensions \cite{Rafelski:1974rh}. From a conceptual point of view, possibly also relevant to semiclassical gravity, it would be interesting to investigate the domain of validity of the semiclassical approximation, \textit{i.e.}, the neglect of quantum fluctuations of the electric field.

\appendix
\section{Convergence of the self-consistent fields}
\label{sec:convergence}

\begin{figure}[t]
    \begin{subfigure}{0.45\textwidth}
        \centering
        \includegraphics[width=\textwidth]{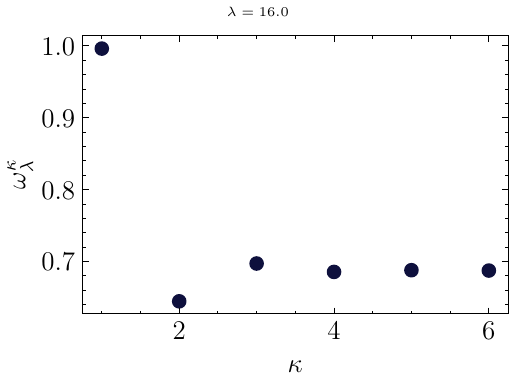}
    \end{subfigure}
  \begin{subfigure}{0.45\textwidth}
        \centering
        \includegraphics[width=\textwidth]{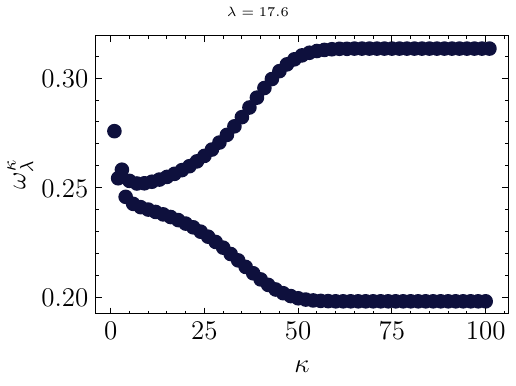}
    \end{subfigure}
        \caption{$\omega_1$ of the massless scalar field with DBC for different values of $\lambda$ as a function of the iteration $\kappa$.}        \label{fig:convergence}

\end{figure}

The iterative procedure either
    converges,
    oscillates, or
    breaks down. 
    
We discuss the convergence of the iterative procedure described in Sec.~\ref{sec:The-iterative-procedure} 
in terms of the sequence $\{\omega_{1, \lambda}^\kappa\}_{\kappa>0}$ of the frequency of the first mode. Examples of a converging sequence and an oscillating (non‑convergent) sequence are displayed in the top and bottom panels of Fig.~\ref{fig:convergence}, respectively. These cases can be understood from the perspective of fixed‑point theory, where we explicitly see whether the update function $f_{\lambda,c}$ is a contraction (top panel) or not (bottom panel). In the range of $\lambda$ that we were considering, it was always possible to choose the ``damping parameter'' $c$ such that convergence is achieved.


The iterative procedure breaks down when complex $\omega$ appear. This happens if the candidate $\lambda +\Delta \lambda $ in the numerical continuation method is too far away from $\lambda$: the screening due to $\rho_\lambda$  is too weak compared to $A_{0, \lambda+\Delta \lambda}^1$, and the system effectively behaves as in the external field approximation with $\lambda>\lambda_c$.
The step $\Delta \lambda$ is reduced and the iterative procedure is tried again. 


\section*{Acknowledgements}
S.S. acknowledges funding by the Department of Education of the Basque
Government through the IKUR Strategy, through BasQ
(project EMISGALA), as well as by the Agencia Estatal de
Investigación (AEI) through Proyectos de Generación de
Conocimiento PID2022-142308NA-I00 (EXQUSMI).
\printbibliography{}

\end{document}